\documentclass[journal]{IEEEtran}

\usepackage{amsmath,amssymb,amsthm}
\usepackage{cite}
\usepackage{epsfig}
\usepackage[ruled,vlined,lined,linesnumbered]{algorithm2e}
\usepackage{color}

\ifCLASSINFOpdf
\else
\fi
\newcommand{\ALARM}[1]{\textcolor{black}{#1}}
\newcommand{\CONV}[1]{\textcolor{black}{#1}}
\newcommand{\RVF}[1]{\textcolor{black}{#1}}
\begin{document}
%
\title{Improved Iterative Hard- and Soft-Reliability Based Majority-Logic Decoding
  Algorithms\\ for Non-Binary Low-Density Parity-Check Codes}

\IEEEoverridecommandlockouts

\author{
\IEEEauthorblockN{Chenrong Xiong and Zhiyuan Yan}
\thanks{This work was supported in part by NSF under Grants ECCS-0925890 and ECCS-1055877.}


}



%


\maketitle

\begin{abstract}
Non-binary low-density parity-check (LDPC) codes have some advantages over their
binary counterparts, but unfortunately their decoding complexity is a
significant challenge. The iterative hard- and soft-reliability based
majority-logic decoding algorithms are attractive for non-binary LDPC codes,
since they involve only finite field additions and multiplications as well as
integer operations and hence have significantly lower complexity
than other algorithms. In this paper, we propose two improvements to the
majority-logic decoding algorithms. Instead of the accumulation of reliability
information in the existing majority-logic decoding algorithms, our first
improvement is a new reliability information update. The new update not only results in better
error performance and \CONV{fewer iterations on average}, but also further reduces computational complexity.
Since existing majority-logic decoding algorithms tend to have a high error
floor for codes whose parity check matrices have low column weights, our
second improvement is a re-selection scheme, which leads to much lower error
floors, at the expense of more finite field operations and integer operations, by
identifying periodic points, re-selecting intermediate hard decisions, and
changing reliability information. 
\end{abstract}


%
\IEEEpeerreviewmaketitle

\section{Introduction}
\label{sec:intro}

\RVF{Low-density parity-check (LDPC) codes were first developed by Gallager
\cite{gallager1963ldpc} in 1963. They were forgotten until they were
rediscovered in the late 1990s by MacKay and Neal \cite{585036}.
Since then, the academic and industrial
communities have focused on binary LDPC codes, because long binary LDPC codes
can achieve performance approaching
the Shannon limit (see, for example, \cite{capacity2001Richardson}). Hence
binary LDPC codes have been used in various applications, such as digital
television \cite{dvbs2-standard}, Ethernet \cite{1700008}, home networking \cite{G9960-standard},
and Wi-Fi \cite{5307322}. Efficient decoding algorithms, encoder
implementations, and decoder implementations of
binary LDPC codes (see, for example, \cite{4915776, 1275678, 5975253, 5475309,
  4602535, 4518325, 5351656, 5605921}) have received significant attentions.}

\RVF{In 1998, the study of Davey and MacKay \cite{davey1998nonbinaryldpc} showed
  that non-binary LDPC codes over GF($q$) ($q>2$) perform better than their binary counterparts for moderate code lengths. Moreover, non-binary LDPC codes
also outperform binary LDPC codes on channels with bursty errors and
high-order modulation schemes \cite{QLDPC2003Song}. These advantages have
motivated a steady stream of work on code designs \cite{ 4840630, Lingqi2008ConstructionNBLDPC, bozhou2009qcnbldpc}, decoding
algorithms \cite{5946693, davey1998nonbinaryldpc, evaluation1999mackay, Barnault2003fftnbldpc, QLDPC2003Song,
  minsum2004wymeersch, declercq2007emsnldpc, savin2008minmax, Chaoyu2010mpa}, and
decoder implementations \cite{5946358,  6363196, Xinmiao2011EIHRB} for non-binary LDPC
codes. Davey and MacKay \cite{davey1998nonbinaryldpc} first used belief
propagation (BP) to decode non-binary LDPC codes. By applying the fast
Fourier transform (FFT) of probabilities to the BP algorithm, they also proposed a fast Fourier transform
(FFT) based $q$-ary sum-product algorithm (SPA), called FFT-QSPA
\cite{evaluation1999mackay}. The FFT-QSPA was further improved by Barnault and
Declercq \cite{Barnault2003fftnbldpc}. Song and Cruz proposed a logarithm domain
FFT-BP algorithm\cite{QLDPC2003Song}. The Min-Sum
algorithm was applied to non-binary LDPC codes by Wymeersch
\textit{et al}. \cite{minsum2004wymeersch}. Then Declercq and Fossorier
\cite{declercq2007emsnldpc} proposed the Extended Min-Sum (EMS) algorithm by using only a limited
number of probabilities in the messages at inputs of check nodes. Savin
\cite{savin2008minmax} proposed the Min-Max algorithm.}

Advantages of non-binary LDPC codes come at the expense of significantly
higher decoding complexity than their binary counterparts. Since 
complexity of decoding non-binary LDPC codes is a key challenge, the
iterative hard- and soft-reliability based majority-logic decoding, referred to
as IHRB-MLGD and ISRB-MLGD\footnote{When there is no ambiguity, MLGD is omitted when referring to majority-logic decoding algorithms for brevity.}, 
respectively, algorithms \cite{Chaoyu2010mpa} are particularly attractive. Based
on the one-step majority logic decoding, these majority-logic decoding
algorithms represent reliability information with finite field elements and integers, and hence involve only
finite field additions (FAs) and finite field multiplications (FMs) as well as integer
additions (IAs), integer comparisons (ICs), integer multiplications (IMs) and
integer divisions (IDs). As a result, they require much lower computational
complexities at the expense of moderate error performance degradation. For
instance, while the error performance of the ISRB algorithm is within 1 dB
of that of FFT-QSPA \cite{Barnault2003fftnbldpc}, its complexity is only
a small fraction of that of the latter \cite{Chaoyu2010mpa}. With a performance
loss of 1 dB, the IHRB algorithm has even lower complexity than the ISRB
algorithm \cite{Chaoyu2010mpa}. Based on the IHRB algorithm, Zhang \textit{et al.}\
\cite{Xinmiao2011EIHRB} proposed an enhanced IHRB-MLGD (EIHRB) algorithm by
introducing the soft-reliability initialization and re-computing the extrinsic
information. The EIHRB algorithm has a similar complexity to that of the 
IHRB algorithm, but its error performance approaches that of the ISRB
algorithm. The majority-logic decoding algorithms are particularly effective for
LDPC codes constructed based on finite geometries and finite fields
\cite{Lingqi2008ConstructionNBLDPC, bozhou2009qcnbldpc}.

The main contributions of this paper are two improvements to the majority-logic decoding algorithms.
\begin{itemize}
\item The first improvement is a new reliability information update, instead of the
  accumulation of reliability information used in existing
  majority-logic decoding algorithms. 
\item Since existing majority-logic decoding algorithms tend to have a high
  error floor for codes whose parity check matrices have small column weights,
  our second improvement is a re-selection scheme, which lowers
  error floors at the expense of more finite field operations and integer
  operations by identifying periodic points, re-selecting intermediate hard
  decisions, and changing reliability information.
\end{itemize}

In the ISRB and IHRB algorithm,
the reliability information includes all check-to-variable (c-to-v) messages of previous
iterations. 
The new reliability information update proposed in this paper excludes the c-to-v messages
of previous iterations. It not only results in better error performance and
\CONV{fewer iterations on average}, but also greatly reduces computational complexities of
\emph{all} existing majority-logic decoding algorithms. For instance, when
applied to the ISRB majority-logic decoding algorithm, the new reliability
information update results in a $0.15$ dB coding gain and \CONV{reduces required
  number of iterations} by 10\% at 4.7 dB for a $(16, 16)$-regular $(255, 175)$
cyclic LDPC code over GF($2^8$) constructed with the method as describe in
\cite[Example 4]{Lingqi2008ConstructionNBLDPC}. Also, at a block error rate (BLER) of
$10^{-4}$, the coding gain over the EIHRB algorithm is about $0.07$ dB. At
the SNR of 4.7 dB, \CONV{the average number of iterations is reduced by} about 25\%. Furthermore, with the
new reliability information update, the improved algorithms require
significantly fewer IAs and ICs than the ISRB and EIHRB
algorithms. Finally, the existing majority-logic decoding algorithms are based
on the accumulation of reliability information, and hence the numerical range
of the reliability information increases with iterations. In contrast, the
proposed reliability information update results in a
fixed numerical range and thus simplifies hardware implementations. \RVF{Our new
  reliability update has been presented in part in
\cite{6190138}. By applying both the layered scheduling and our first
improvement to the IHRB algorithm, we proposed a layered improved IHRB decoder
with a high throughput in \cite{6572104}. Because the architecture design of
non-binary LDPC decoders is beyond the scope of this paper, we will not discuss
the layered improved IHRB decoder henceforth.} 

In the literature, to analyze the error floor of binary LDPC codes, some notions
based on graphical structures have been introduced, such as stopping
sets \cite{1003839}, trapping sets \cite{2003errorfloor} and absorbing sets
\cite{zhengya2008errorfloor}. Unfortunately, trying to lower the error floor
based on graphical structures usually incurs very high
complexity. Also, some approaches for binary LDPC codes cannot be readily
adapted to non-binary ones.
For instance, a selective biasing postprocessing algorithm is proposed in
\cite{zhengya2008errorfloor} to lower the error floors of binary LDPC 
codes based on the relaxed graphical structure of absorbing sets. However, for
non-binary LDPC codes, trapping sets are difficult to identify because
they involve not only the graph topology but
also values of non-zero entries of parity-check matrices \cite{6325207}. Moreover, the biasing rule between two elements for binary LDPC codes cannot be applied to
non-binary codes directly, because there are more than two elements in a
non-binary finite field.

In this paper, for the majority-logic decoding
algorithms, we propose a re-selection scheme based on periodic points to
lower the error floors. The re-selection scheme is not a postprocessing
algorithm and can be integrated into the regular iteration procedure easily. For instance, for an
(837, 726) non-binary quasi-cyclic LDPC code over GF($2^5$) constructed with
the method in \cite{bozhou2009qcnbldpc} with a column weight of four, the
EIHRB algorithm has a BLER floor around $10^{-3}$, while the hard-reliability
based algorithm with the new reliability information update and the re-selection
scheme achieves a BLER floor below $10^{-5}$. Although this re-selection scheme
requires additional computation, it is used only when existing majority-logic
decoding algorithms have a high error floor.

The rest of our paper is organized as follows. Section~\ref{sec:review} reviews
existing majority decoding algorithms. Section~\ref{sec:algorithm} proposes 
the two improvements. In
Section~\ref{sec:sim_result}, the two improvements are applied to existing majority
decoding algorithms to illustrate their advantages in error performance and
\CONV{average numbers of iterations}. Section~\ref{sec:complexity} discusses the reduction in the
computational complexities due to the two improvements. Some conclusions are
given in Section \ref{sec:conclusion}.

\section{Existing Majority Decoding Algorithms}
\label{sec:review}

A regular LDPC code $\mathcal{C}$ of length $N$ over a finite field GF$(2^r)$ is
the null space of an $M \times N$ sparse parity check
matrix $\mathbf{H}$ over GF($2^r$). $\mathbf{H}$ has constant column and row
weights of $\gamma$ and $\rho$, respectively. Let
$\mathbf{h}_0,\mathbf{h}_1,\cdots,\mathbf{h}_{M-1}$ denote the
rows of $\mathbf{H}$, where $\mathbf{h}_i=(h_{i,0},h_{i,1},\cdots,h_{i,N-1})$ for
$0 \leq i < M$. \ALARM{Let $(a_{l,0},a_{l,1},\cdots,a_{l,r-1})$ be the binary
representation of $a_l\in \text{GF}(2^r)$, for $0 \leq l < 2^r$.} Suppose a codeword
$\mathbf{x}=(x_0,x_1,\cdots,x_{N-1})$ is transmitted. Since $x_i \in \text{GF}(2^r)$ can be represented by an $r$-tuple $(x_{i,0},x_{i,1},\cdots,x_{i,r-1})$ over
GF($2$) for $0 \leq i < N$, an $Nr$-tuple over GF($2$) is transmitted for each
codeword. Assume the
BPSK modulation is used: ``0" is mapped to +1 and ``1" to -1. Let $\mathbf{y}=(y_0,y_1,\cdots,y_{N-1})$ represent the received word, and $\mathbf{z}=(z_0,z_1,\cdots,z_{N-1})$ and
$\mathbf{q}=(q_0,q_1,\cdots,q_{N-1})$ represent the hard decision and
quantization, respectively, of the received word. Let $\mathcal{N}(i) = \{j:h_{i,j}\neq 0, 0
\leq j < N\}$ for $0 \leq i < M$ and $\mathcal{M}(j)=\{i:h_{i,j}\neq 0, 0 \leq i < M\}$ for
$0 \leq j < N$. $I_{\rm{max}}$ represents the maximal iteration number.

\subsection{ISRB algorithm}
\label{sec:rev-ISRB}
\begin{algorithm}
\caption{ISRB algorithm \cite{Chaoyu2010mpa}}
\label{alg:ISRB}
\LinesNumbered
\tcc{---------Initialization----------}
\For{$j=0:(N-1)$}{
$z_j^{(0)} = z_j$\;
\For{$l=0:(2^{r}-1)$}{
 $\varphi_{j,l}=\sum_{t=0}^{r-1} (1-2a_{l,t})q_{j,t}$\;
$R_{j,l}^{(0)}=\lambda \varphi_{j,l}$\;
}
}
\For{$i=0:(M-1)$}{
\For{$j \in \mathcal{N}(i)$}{
$\phi_{i,j}=\min_{t\in \mathcal{N}(i) \setminus \{j\}} \max_l \varphi_{t,l}$\;
}
}
\tcc{-----------Iteration------------}
\For{$k=0: I_{\rm{max}}$}{
 $\mathbf{s}^{(k)}=\mathbf{H} \cdot (\mathbf{z}^{(k)})^{T}$\;
\lIf{$\mathbf{s}^{(k)}==0$}{
\Return{$\mathbf{z}^{(k)}$}
}
\lElseIf{$k==I_{\rm{max}}$}{
\Return{$\text{Failure}$}
}
\uElse{\label{r4}
\For{$j=0:(N-1)$}{\label{r5}
\For{$l=0:(2^r-1)$}{
$\psi_{j,l}^{(k)}=0$\;
}
\For{$i \in \mathcal{M}(j)$}{
$\sigma_{i,j}^{(k)}=h_{i,j}^{-1}\sum_{t\in \mathcal{N}(i) \setminus
  \{j\}}h_{i,t}z_{t}^{(k)}$\label{r6}\;
\ALARM{
\For{$l=0:(2^r-1)$}{
\lIf{$\sigma_{i,j}^{(k)}==a_l$}{
$\psi_{j,l}^{(k)}=\psi_{j,l}^{(k)}+\phi_{i,j}$\label{r2:ISRB}
}
}
}
}
}
}
\For{$j=0:(N-1)$}{
\For{$l=0:(2^r-1)$}{
$R_{j,l}^{(k+1)}=R_{j,l}^{(k)}+\psi_{j,l}^{(k)}$\label{r1}\;
}
$z_j^{(k+1)}=\arg_{a_l} \max R_{j,l}^{(k+1)}$\label{r3}\;
}
}
\end{algorithm}

The ISRB algorithm \cite{Chaoyu2010mpa} is described in
Alg. \ref{alg:ISRB}, where $\lambda$ is a parameter to improve the error
performance. \RVF{$\mathbf{s}^{(k)}$ is the syndrome vector corresponding to
  $\mathbf{z}^{(k)}$, $\varphi_{j,l}$ a channel reliability of the
  $j$-th received symbol being $a_l$. $\phi_{i,j}$ and $\sigma_{i,j}^{(k)}$ are the extrinsic
  weighting coefficient and the extrinsic check sum of the $k$-th iteration, respectively,
  from check node $i$ to variable node $j$. $\psi_{j,l}^{(k)}$ is the extrinsic
  reliability of the $j$-th received symbol being $a_l$ in the $k$-th iteration.} 

In the ISRB algorithm, line~\ref{r1} is an accumulation
operation. Hence, the reliability $R_{j,l}^{(k)}$ is a
non-decreasing function of $k$ as $\psi_{j,l}^{(k)}$ is
non-negative. To perform the ISRB algorithm correctly,
$R_{j,l}^{(k)}$ must be kept from numerical saturation based on two
methods. One is to use a very large numerical range for $R_{j,l}^{(k)}$, and the other is to carry out the following clipping operation
\cite{Chaoyu2010mpa}:

\begin{equation}\label{clipping}
R_{j,l}^{(k)}=\left\{
\begin{array}{ll}
-\eta & \text{if $R_{j,l}^{(k)}< R_{j,max}^{(k)}-2\eta$} \\
R_{j,l}^{(k)}-R_{j,max}^{(k)}+\eta & \text{otherwise} \\
\end{array}  \right.
\end{equation}

Here, $R_{j,max}^{(k)} \triangleq \max_l (R_{j,l}^{(k)})$ and \ALARM{$\eta$
is the predefined maximal value of $R_{j,l}^{(k)}$ after the
clipping operation.}


\subsection{IHRB algorithm}
\label{sec:rev-IHRB}
When the soft-reliability information of the received word is not available to
the decoder, the IHRB algorithm \cite{Chaoyu2010mpa} can be used.
 The iteration procedure of the IHRB algorithm is the same as that of the ISRB
algorithm, but the IHRB algorithm has a different initialization step, described in
Alg.~\ref{alg:IHRB-INIT}, where $\lambda_h$ is a parameter to improve the
error performance.

\begin{algorithm}
\caption{Initialization of the IHRB algorithm \cite{Chaoyu2010mpa}}
\label{alg:IHRB-INIT}
\For{$j=0:(N-1)$}{
$z_j^{(0)} = z_j$\;
\For{$l=0:(2^{r}-1)$}{
\lIf{$(a_l==z_j)$}{
$R_{j,l}^{(0)}=\lambda_h$
}
\lElse{
$R_{j,l}^{(0)}=0$
}
}
}
\For{$i=0:(M-1)$}{
\For{$j \in \mathcal{N}(i)$}{
$\phi_{i,j}=1$\;
}
}
\end{algorithm}

\subsection{EIHRB algorithm}
\label{sec:rev-EIHRB}
The EIHRB algorithm \cite{Xinmiao2011EIHRB}, described by Alg.~\ref{alg:EIHRB},
was devised based on the IHRB algorithm by introducing a soft-reliability
initialization and recalculating the extrinsic information. $c_1$ and $c_2$ are
two parameters to improve the error performance.

\begin{algorithm}
\caption{EIHRB algorithm \cite{Xinmiao2011EIHRB}}
\label{alg:EIHRB}
\LinesNumbered
\tcc{---------Initialization----------}
\For{$j=0:(N-1)$}{
$z_j^{(0)} = z_j$\;
$z_{i.j}^{(0)}=z_j$\;
\For{$l=0:(2^{r}-1)$}{
 $\varphi_{j,l}=\sum_{t=0}^{r-1} (1-2a_{l,t})q_{j,t}$\;
$R_{j,l}^{(0)}=\max(\lfloor \varphi_{j,l} / c_1 \rfloor+c_2 - \max_l(\lfloor\varphi_{j,l} / c_1\rfloor),0)$\label{r8}\;
}
}
\tcc{-----------Iteration------------}
\For{$k=0: I_{\rm{max}}$}{
 $\mathbf{s}^{(k)}=\mathbf{H} \cdot (\mathbf{z}^{(k)})^{T}$\;
\lIf{$\mathbf{s}^{(k)}==0$}{
\Return{$\mathbf{z}^{(k)}$}
}
\lElseIf{$k==I_{\rm{max}}$}{
\Return{$\text{Failure}$}
}
\uElse{
\For{$i=0:(M-1)$}{
\For{$j \in \mathcal{N}(i)$}{
$\sigma_{i,j}^{(k)}=h_{i,j}^{-1}\sum_{t\in \mathcal{N}(i) \setminus \{j\}}h_{i,t}z_{i,t}^{(k)}$\;
\ALARM{
\For{$l=0:(2^r-1)$}{
\lIf{$\sigma_{i,j}^{(k)}==a_l$}{
$R_{j,l}^{(k)}=R_{j,l}^{(k)}+1$\label{r2:EIHRB}
}
$R_{j,l}^{(k+1)}=R_{j,l}^{(k)}$\;
}
}
}
}
}
\For{$j=0:(N-1)$}{\label{a2_b}
$R_{j}^{(k+1)\text{m}}=\max_l R_{j,l}^{(k+1)}$\;
$z_j^{(k+1)}=\text{field element of }R_{j}^{(k+1)\text{m}}$\;
$R_{j}^{(k+1)\text{m2}}=\text{second largest among } R_{j,l}^{(k+1)}$\;
$z_j^{\prime (k+1)}=\text{field element of }R_{j}^{(k+1)\text{m2}}$\;
\For{$i \in \mathcal{M}(j)$}{
\lIf{$(\sigma_{i,j}^{(k)}==z_j^{(k+1)})\&(R_{j}^{(k+1)\text{m}} \leq R_{j}^{(k+1)\text{m2}})+1$}{
$z_{i,j}^{(k+1)}=z_j^{\prime (k+1)}$
}
\lElse{
$z_{i,j}^{(k+1)}=z_j^{(k+1)}$\label{a2_e}
}
}
}
}
\end{algorithm}

If the soft-reliability information of the received symbol is available, the
EIHRB algorithm achieves a better error performance than the IHRB
algorithm. Therefore, we focus on the EIHRB algorithm and do not consider
the IHRB algorithm further.

\section{Two Improvements}

\label{sec:algorithm}
\subsection{New Reliability Information Update}
\label{subsec:new_rlb_upd}
The reliability information update of line~\ref{r1} of Alg.~\ref{alg:ISRB} can be written as:
\begin{equation}\label{rlb_update_re}
\begin{split}
R_{j,l}^{(k+1)} & = R_{j,l}^{(k)} + \psi_{j,l}^{(k)} \\
& = R_{j,l}^{(0)} + \sum_{t=0}^{k} \psi_{j,l}^{(t)} 
\end{split}
\end{equation}

For both the ISRB and IHRB algorithms, the reliability information of the $k$-th
iteration, $R_{j,l}^{(k)}$, includes all
check-to-variable (c-to-v) messages of previous iterations. This conflicts with
the extrinsic information principle. In the EIHRB algorithm, lines \ref{a2_b} to \ref{a2_e} of
Alg.~\ref{alg:EIHRB} are used to recalculate the extrinsic information.

We propose a new reliability information update to exclude the c-to-v messages of previous
iterations. In our new reliability information update, only the channel information
$\varphi_{j,l}$ and $\psi_{j,l}^{(k)}$ of the current iteration are used to
compute the reliability information $R_{j,l}^{(k+1)}$. Our new reliability
information update is

\begin{equation}\label{new_rlb_update_1}
R_{j,l}^{(k+1)}=\xi_1\varphi_{j,l} + \xi_2\psi_{j,l}^{(k)},
\end{equation}
where $\xi_1$ and $\xi_2$ are two parameters to improve the error performance.

Eq.~\eqref{new_rlb_update_1} is used to replace the reliability information update of
line~\ref{r1} of Alg.~\ref{alg:ISRB}, and consequently the new algorithm is
called the IISRB algorithm.  

\ALARM{
To reduce complexity of the IISRB algorithm, we change the
initialization as follows. 
For the ISRB algorithm,
$\phi_{i,j}$ and $\varphi_{j,l}$ are calculated in the initialization. Hence, for the
IISRB algorithm, we calculate $\xi_1 \varphi_{i,j}$ and $\xi_2 \phi_{i,j}$ in the
initialization as well. This helps to reduce the complexity of each iteration. The
IISRB algorithm is presented in Alg.~\ref{alg:init_IISRB}.
}

\begin{algorithm}
\caption{IISRB algorithm}
\label{alg:init_IISRB}
\ALARM{
\tcc{---------Initialization----------}
\For{$j=0:(N-1)$}{
$z_j^{(0)} = z_j$\;
\For{$l=0:(2^{r}-1)$}{
 $\varphi_{j,l}^{\prime}=\sum_{t=0}^{r-1} (1-2a_{l,t})q_{j,t}$\;
$\varphi_{j,l}=\xi_1\varphi_{j,l}^{\prime}$\;
}
}
\For{$i=0:(M-1)$}{
\For{$j \in \mathcal{N}(i)$}{
$\phi_{i,j}=\xi_2\min_{t\in \mathcal{N}(i) \setminus \{j\}} \max_l \varphi_{t,l}^{\prime}$\;
}
}
\tcc{-----------Iteration------------}
\For{$k=0: I_{\rm{max}}$}{
 $\mathbf{s}^{(k)}=\mathbf{H} \cdot (\mathbf{z}^{(k)})^{T}$\;
\lIf{$\mathbf{s}^{(k)}==0$}{
\Return{$\mathbf{z}^{(k)}$}
}
\lElseIf{$k==I_{\rm{max}}$}{
\Return{$\text{Failure}$}
}
\uElse{\label{r4:IISRB}
\For{$j=0:(N-1)$}{\label{r5:IISRB}
\For{$l=0:(2^r-1)$}{
$\psi_{j,l}^{(k)}=0$\;
}
\For{$i \in \mathcal{M}(j)$}{
$\sigma_{i,j}^{(k)}=h_{i,j}^{-1}\sum_{t\in \mathcal{N}(i) \setminus \{j\}}h_{i,t}z_{t}^{(k)}$\label{r6:IISRB}\;
\For{$l=0:(2^r-1)$}{
\lIf{$\sigma_{i,j}^{(k)}==a_l$}{
$\psi_{j,l}^{(k)}=\psi_{j,l}^{(k)}+\phi_{i,j}$\label{r2:IISRB}
}
}
}
}
}
\For{$j=0:(N-1)$}{
\For{$l=0:(2^r-1)$}{
$R_{j,l}^{(k+1)}=\varphi_{j,l}+\psi_{j,l}^{(k)}$\label{r1:IISRB}\;
}
$z_j^{(k+1)}=\arg_{a_l} \max R_{j,l}^{(k+1)}$\label{r3:IISRB}\;
}
}
}
\end{algorithm}

A new reliability information is also applied to the EIHRB
algorithm. The reliability information update in line~\ref{r2:EIHRB} of
Alg.~\ref{alg:EIHRB} is replaced with 

\begin{equation}
\label{HRB-update}
R_{j,l}^{(k)}=R_{j,l}^{(k)} + c_3,
\end{equation}
where, $c_3$ is a parameter to improve the error performance. Meanwhile, at the
beginning of each iteration, $R_{j,l}^{(k)}$ is initialized as $R_{j,l}^{(0)}$
which is already scaled in line~\ref{r8} by a parameter $c_1$. Furthermore, to be consistent with the IISRB algorithm,
$z_{i,j}^{(k+1)}=z_j^{(k+1)}$. Finally, lines \ref{a2_b} to \ref{a2_e} of
Alg.~\ref{alg:EIHRB} are not needed any more. The new algorithm derived from the
EIHRB algorithm with the four modifications above is referred to as the IEIHRB
algorithm.


\subsection{Re-selection Scheme}
\label{subsec:RS}
%

Furthermore, we observe that the error floor of the ISRB algorithm becomes
higher, as the column weight of the parity check matrix decreases. The
IISRB algorithm suffers the same problem. 

\begin{figure*}[htbp]
\centering
\includegraphics[width=14cm]{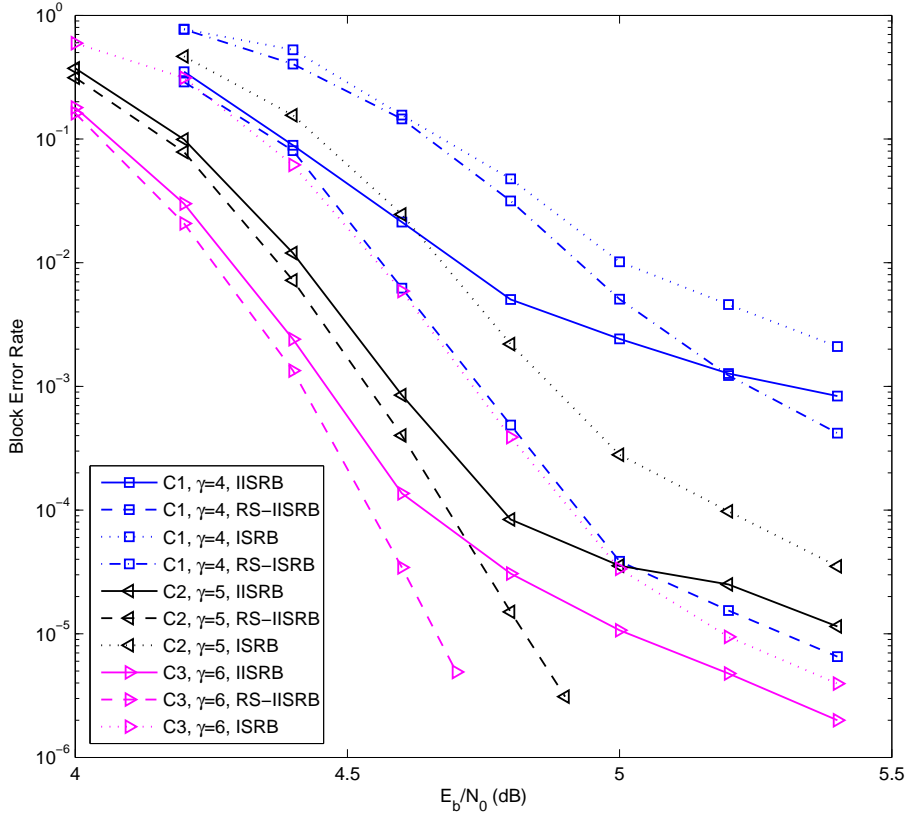}
\caption{Block error rates of the soft-reliability based algorithms for different codes with different column
  weights over the AWGN channel when $I_{\rm{max}}=50$ and the modulation
  scheme is BPSK}
\label{fig:per_ef_cw}
\end{figure*}

In Fig.~\ref{fig:per_ef_cw}, C1 is an (837,
726) LDPC code over GF($2^5$) with a column weight of four, C2 an (806, 680) LDPC code over GF($2^5$) with a column
weight of five, C3 a (775, 634) LDPC code over
GF($2^5$) with a column weight of six. All three codes are
constructed based on Reed--Solomen codes with two information symbols
\cite{bozhou2009qcnbldpc}. The error floor of BLER performance becomes lower as the
column weight of the parity check matrix increases. Hence, the column weight of
the parity check matrix is one key factor for the error floor. 

We propose a re-selection scheme to address this problem. 

To simplify the discussion, here we focus on the IISRB algorithm. 
Our simulation results show that the re-selection scheme
also applies to the ISRB, EIHRB and IEIHRB algorithms.

To analyze the error floor, the concept of periodic points is introduced. 
Given an endomorphism $f:Z \rightarrow Z$, a
point $\mathbf{z}$ in $Z$ is called a \emph{periodic point with a period of $i$} if there
exists a smallest positive integer $i$ so that $f^{(i)}(\mathbf{z})=\mathbf{z}$,
where $f^{(i)}=f(f^{(i-1)}(\mathbf{z}))$. 

An iteration of the IISRB algorithm can be considered a function $f$. The $k$-th iteration of the IISRB algorithm is
$\mathbf{z}^{(k)}=f(\mathbf{z}^{(k-1)})=f^{(2)}(\mathbf{z}^{(k-2)})=\cdots=f^{(k)}(\mathbf{z}^{(0)})$,
and if $\mathbf{s}^{(k)} \neq 0$ and $\mathbf{z}^{(k)}=\mathbf{z}^{(k-i)}$ for
$0 < i\leq k$, the decoding algorithm results in a periodic point with a period
of $i$. 
Our algorithm focuses on only the periodic points with a period of up to two for two
reasons. First, our simulation results show that the BLERs are caused mainly by
periodic points with periods one and two. Second, to identify the existence of
a periodic point with a period of greater than two needs more memory to keep
track of the hard decisions of the previous iterations.

%

\begin{figure}[htbp]
\centering
\includegraphics[width=9cm]{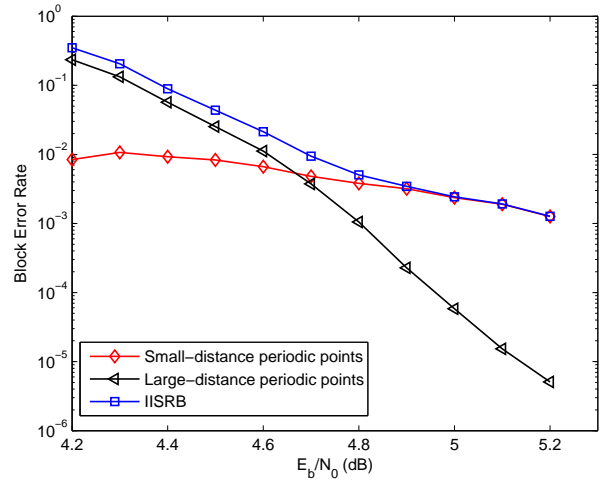}
\caption{BLERs of the small-distance and large-distance
  periodic points for the IISRB algorithm to decode the (837, 726) code over the
  AWGN channel when the modulation scheme is BPSK}
\label{fig:837-iisrb-tp}
\end{figure}

If the Hamming distance between a periodic point and its corresponding
transmitted codeword is less than $\theta$, the periodic point 
is called a small-distance periodic point. Otherwise, it is called a
large-distance periodic point. Fig.~\ref{fig:837-iisrb-tp} compares the BLERs of
the IISRB algorithm with those caused by the large-distance and small-distance
periodic points for the (837, 726)
code when $\theta = 8$. 
For low SNRs, the overall BLER is dominated by those caused by large-distance
periodic points. The sum of the BLERs due to the large-distance and the
small-distance periodic points is less than the overall BLER, because periodic
points with a period greater than two also cause some BLERs. 
For high SNRs, the BLER caused by the small-distance periodic points dominates the
total BLER.
A similar trend for the ISRB algorithm was observed as well. Hence, for the
ISRB and IISRB algorithms, the error floor is mainly caused by the small-distance periodic points. In order to lower the error
floor of IISRB algorithm, the BLER caused by the small-distance periodic
points should be reduced.

Consider the hard decision process of line \ref{r3} of Alg.~\ref{alg:ISRB}. If
the most likely decision is wrong, the second most likely decision is supposed
to be the best   choice to be decoded. The smaller the difference between the
maximal reliability information and the second maximal reliability information,
the greater the probability that the most likely decision is wrong.

Based on this intuition, when a periodic point is detected, the re-selection
scheme tries to help the decoder get away from the periodic point by using the
second most likely decision. The re-selection scheme consists of two steps. The
first step is to identify the existence of a periodic point when the syndrome
vector is a non-zero vector. The second step is to identify positions of
erroneous symbols. A set is defined to include variable nodes adjacent to
unsatisfied check nodes. This set contains some erroneous symbols. Then,
among the variable nodes belonging to the set, the position of a
variable node which has the smallest difference between its maximal reliability
information and second maximal reliability information can
be identified. If there are multiple variable nodes having the smallest
difference, the first one is selected. Assume the index of this position is
$\rm{rs\_n}$. Let ${\rm{us\_c}}_j$ represent the number of unsatisfied check
nodes connected with the $j$-th variable node for $0\leq j < N$. The most likely
decision $z_{\rm{rs\_n}}^{(k)}$ is replaced by the
second most likely decision $\tilde{z}_{\rm{rs\_n}}^{(k)}$. Meanwhile,
$\varphi_{{\rm{rs\_n}},z_{\rm{rs\_n}}^{(k)}}$ is reduced by a preset offset
$\zeta$ and $\varphi_{{\rm{rs\_n}},\tilde{z}_{\rm{rs\_n}}^{(k)}}$ is added by
the same preset offset. The detailed re-selection scheme is described in
Alg.~\ref{alg:RS-Scheme}. \ALARM{Here, $s_i^{(k)}$ is the $i$-th value of the syndrome
vector $\mathbf{s}^{(k)}$.}

\begin{algorithm}
\caption{Re-selection scheme}
\label{alg:RS-Scheme}
\LinesNumbered
\For{$j=0:(N-1)$}{
$\tilde{z}_j^{(k)} =\arg_{a_l\in GF(2^r) \setminus \{z_j^{(k)}\}} \max R_{j,l}^{(k)}$\;
}
\uIf{$(\mathbf{z}^{(k-1)}==\mathbf{z}^{(k)})$ \bf{or} $(\mathbf{z}^{(k-2)}==\mathbf{z}^{(k)})$\label{r7}}{
$\rm{dif\_R}$$=R_{0,z_0^{(k)}}^{(k)}$\;}
\For{$j=0:(N-1)$}{
${\rm{us\_c}}_j=0$\;
\For{$i \in \mathcal{M}(j)$}{
\lIf{$(s_i^{(k)} > 0)$}{${\rm{us\_c}}_j++$}
}
\uIf{$({\rm{us\_c}}_j>0)$ \bf{and} $((R_{j,z_j^{(k)}}^{(k)}-R_{j,\tilde{z}_j^{(k)}}^{(k)})<{\rm{dif\_R}})$}{
${\rm{dif\_R}} = (R_{j,z_j^{(k)}}^{(k)} -R_{j,\tilde{z}_j^{(k)}}^{(k)})$\;
${\rm{rs\_n}}= j$\;
}
}
$\varphi_{{\rm{rs\_n}},z_{{\rm{rs\_n}}}^{(k)}}=\varphi_{{\rm{rs\_n}},z_{{\rm{rs\_n}}}^{(k)}}-\zeta$\;
$\varphi_{\rm{rs\_n},\tilde{z}_{\rm{rs\_n}}^{(k)}}=\varphi_{\rm{rs\_n},\tilde{z}_{\rm{rs\_n}}^{(k)}}+\zeta$\;
$z_{\rm{rs\_n}}^{(k)} = \tilde{z}_{\rm{rs\_n}}^{(k)}$\;
\For{$ i \in \mathcal{M}(\rm{rs\_n})$}{
$s_i^{(k)}=\mathbf{h}_i\cdot(\mathbf{z}^{(k)})^T$\;
}
\end{algorithm}

This scheme can be applied to any majority decoding algorithms. For the
ISRB algorithm, this scheme is added between lines~\ref{r4} and
\ref{r5}. Similarly, the re-selection scheme can be inserted at the
corresponding position of other algorithms. ``RS-'' is prefixed in front of the
name of the algorithms to show that an algorithm adopts the re-selection
scheme. For instance, the ISRB algorithm with the re-selection scheme is called
as the RS-ISRB algorithm.  

\begin{figure}[!htbp]
\centering
\includegraphics[width=9cm]{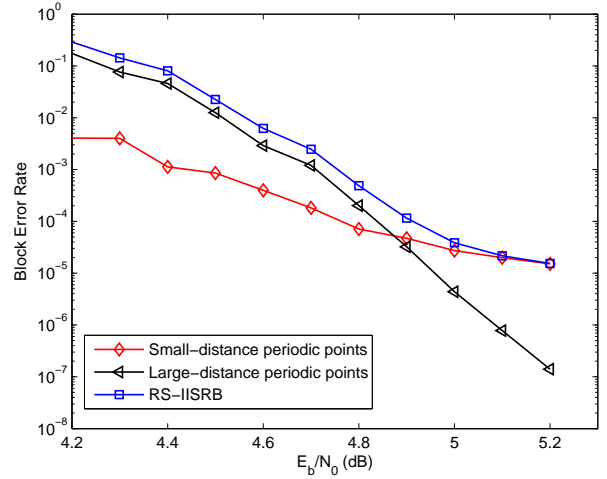}
\caption{BLERs of the small-distance and large-distance
  periodic points for the RS-IISRB algorithm to decode the (837, 726) code over
  the AWGN channel when the modulation scheme is BPSK}
\label{fig:837-rsiisrb-tp}
\end{figure}

Fig.~\ref{fig:837-rsiisrb-tp} shows the BLERs of the RS-IISRB algorithm and
those caused by the
low-distance and high-distance periodic points. Compared with the IISRB algorithm, the BLER caused by the
high-distance periodic points descends to $2\times 10^{-4}$ from $1.2\times
10^{-3}$, and the BLER caused by the low-distance periodic points is reduced to
$7\times 10^{-5}$ from $4\times 10^{-3}$ when SNR is 4.8 dB. Hence, the rs-selection scheme reduces
the occurrences of both the low-distance and high-distance periodic points and
works better on the low-distance periodic points. Even for the RS-IISRB
algorithm, the low-distance periodic points still are the primary reason for the
error floor.

The re-selection scheme helps the decoding algorithm correct
some periodic points. It is likely that the decoding algorithm goes out of a periodic
point temporarily, and goes back to the same periodic point or results in another
periodic point. Therefore, even with the re-selection scheme, the decoding
algorithm still encounters the error floor problem. Moreover, the
re-selection scheme works better on the small-distance periodic points because
in general a small-distance periodic point involves fewer unsatisfied
check nodes than a large-distance periodic point.

\section{Performance Evaluation}
\subsection{Error Performance and \CONV{Average Numbers of Iterations}}
\label{sec:sim_result}


The BPSK modulation scheme, the additive white Gaussian noise (AWGN) channel
with a single-sided power spectral density $N_0$, and a 6-bit uniform
quantization with 64 levels which has an interval length $\Delta=0.0625$ are used in
our numerical simulations. The maximum number of iterations is $50$, i.e., 
$I_{\rm{max}}=50$. Our simulations focus on C1, C2, and C3, whose parity check
matrices have small column weights, as well as a (255,175) cyclic LDPC code over GF($2^8$)
constructed with the method as describe in \cite[Example
4]{Lingqi2008ConstructionNBLDPC}, because it has a large column weight of 16.



\begin{figure}[htbp]
\centering
\includegraphics[width=9cm]{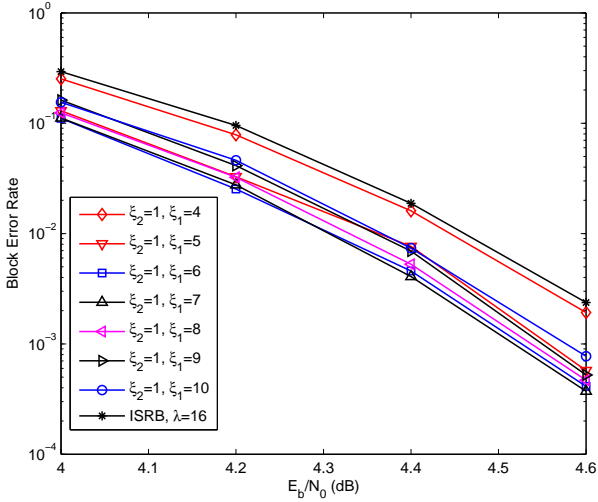}
\caption{The impact of different integer values for $\xi_1$ on the BLER of the IISRB
  algorithm for the (255, 175) code when $I_{\rm{max}}=50$ and the modulation
  scheme is BPSK over the AWGN channel}
\label{fig:255_coef}
\end{figure}

\begin{figure}[htbp]
\centering
\includegraphics[width=9cm]{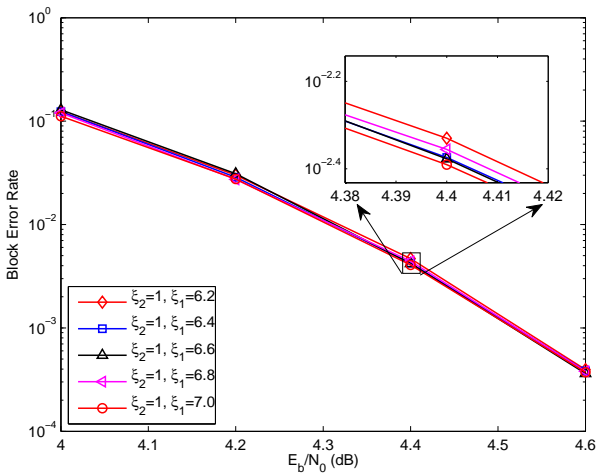}
\caption{The impact of different real values for $\xi_1$ on the BLER of the IISRB
  algorithm to decode the (255, 175) code when $I_{\rm{max}}=50$ and the
  modulation scheme is BPSK over the AWGN channel}
\label{fig:255_coef_flt}
\end{figure}

We first compare the performance of the soft-reliability based algorithms. 
The ISRB, IISRB and RS-IISRB algorithms are used to decode the (255, 175)
code. For the ISRB algorithm, different values of $\lambda=4l$ for
$l=1,2,\cdots,8,$ were tried, and $\lambda=16$ leads to the best performance. For
the new reliability information update, different combinations of $\xi_1$ and
$\xi_2$ were tested. Since they are  weighting factors, we fix $\xi_2=1$ and try
different values for $\xi_1$. As shown in Fig.~\ref{fig:255_coef}, for the (255,
175) code, ($\xi_1=7$, $\xi_2=1$) results in the best error performance. The
real values from 6.2 to 7 with a step size of 0.2 for $\xi_1$ and $\xi_2=1$ were
tested, shown in Fig.~\ref{fig:255_coef_flt}. Performance differences between
different real value coefficients are very small. Henceforth, integer values are
used for $\xi_1$ and $\xi_2$ to reduce complexity.

\begin{figure}[htbp]
\centering
\includegraphics[width=9cm]{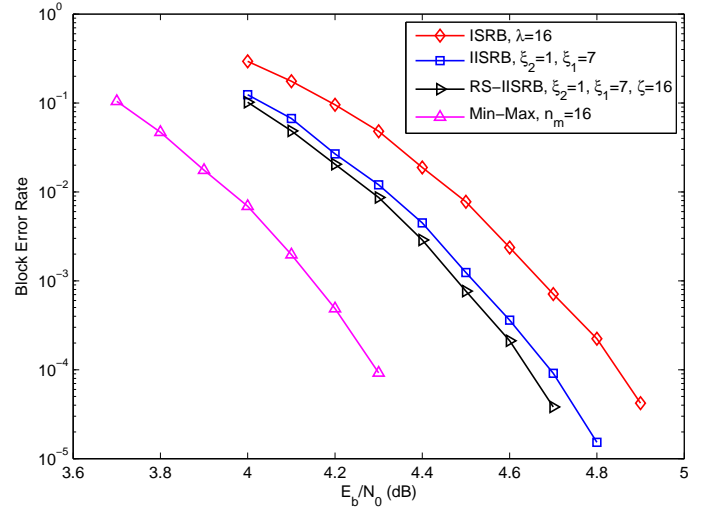}
\caption{Block error rates of the ISRB, IISRB and RS-IISRB algorithms for the
  (255, 175) code when $I_{\rm{max}}=50$ and the modulation scheme is BPSK over the AWGN channel}
\label{fig:BLER_255}
\end{figure}

The BLER curves of the ISRB, IISRB, RS-IISRB
and Min-Max algorithms for the (255,175) code are shown in
Fig.~\ref{fig:BLER_255}. The IISRB algorithm has a 0.15 dB
coding gain versus the ISRB algorithm in this case. The RS-IISRB
algorithm also achieves a slight improvement compared to the
IISRB algorithm and has a performance loss of about 0.4 dB
versus the Min-Max algorithm.

If a total of $T$ iterations is used to decode
$K$ received words, the average number of iterations per received word is
$T/K$.
The average numbers of iterations per received word for the soft-reliability algorithms are compared
in Table~\ref{tab:ave_conv_steps_255}, where $K$ is chosen such that at least
100 erroneous decoded words are observed for each
SNR. Table~\ref{tab:ave_conv_steps_255} shows that both the RS-IISRB
and IISRB algorithms \CONV{require fewer iterations} than the ISRB
algorithm. At 4.7 dB, \CONV{the average number of iterations of the IISRB algorithm is fewer by $10\%$ than that of the ISRB
algorithm.} The advantage of the IISRB and RS-IISRB algorithm is even more
pronounced for low SNRs.

\begin{table}[htbp]
\centering
\caption{Average number of iterations of the Min-Max $(n_m=16)$,
  ISRB, IISRB and RS-IISRB algorithms for the (255, 175)
  code when $I_{\rm{max}}=50$ and the modulation scheme is BPSK over the AWGN channel}
\label{tab:ave_conv_steps_255}
\begin{tabular}{|c|c|c||c|c|}
\hline
$E_b/N_0$ & Min-Max  & ISRB & IISRB & RS-IISRB\\ 
 (dB) &  \cite{savin2008minmax} &  \cite{Chaoyu2010mpa} &  & \\ \hline
4.0 & 2.35 & 18.76 & 11.58 & 11.25 \\ \hline
4.1 & 1.91 & 13.25 &  8.17 & 7.85 \\ \hline
4.2 & 1.60 &  9.10 &  5.78 & 5.71 \\ \hline
4.3 & 1.36 &  6.46 &  4.46 & 4.41 \\ \hline
4.4 & N/A  &  4.59 &  3.59 & 3.59 \\ \hline
4.5 & N/A  &  3.68 &  3.06 & 3.05 \\ \hline
4.6 & N/A  &  3.10 &  2.71 & 2.70 \\ \hline
4.7 & N/A  &  2.74 &  2.46 & 2.46 \\ \hline
4.8 & N/A  &  2.50 &  2.27 & 2.28 \\ \hline
\end{tabular}
\end{table}

\begin{figure}[htbp]
\centering
\includegraphics[width=9cm]{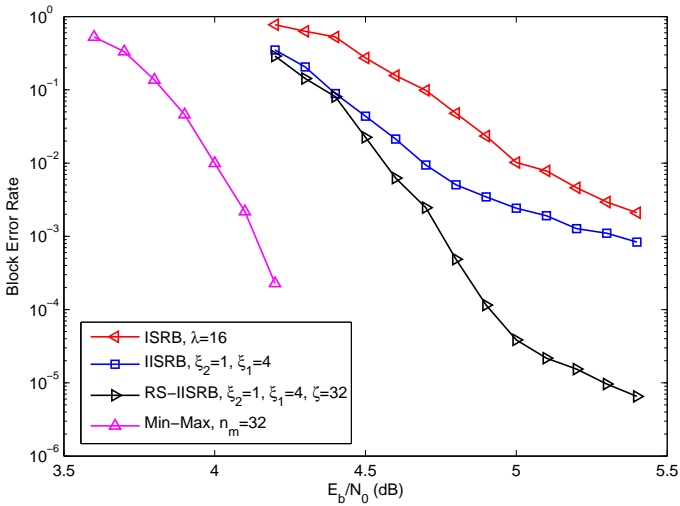}
\caption{Block error rates of the ISRB, IISRB and RS-IISRB
  algorithms for the (837, 726) code when $I_{\rm{max}}=50$ and the modulation
  scheme is BPSK over the AWGN channel}
\label{fig:BLER_837}
\end{figure}


The ISRB, IISRB and RS-IISRB algorithms are also used to
decode the (837, 726) code.
The best BLER performance of the IISRB algorithm is achieved when $\xi_1=4$ and
$\xi_2=1$. Fig.~\ref{fig:BLER_837} compares the BLERs of the ISRB, IISRB, RS-IISRB and Min-Max
algorithms for this code. The IISRB algorithm has a 0.2 dB coding gain versus the ISRB
algorithm, but both algorithms show an error floor around $10^{-3}$. Compared with these two algorithms, for low SNRs the RS-IISRB
algorithm shows a slight improvement, and
for high SNRs the RS-IISRB algorithm lowers the error floor to below $10^{-5}$ and
has a performance loss of only 0.6 dB versus the Min-Max
algorithm at the BLER of $10^{-3}$. 

The average numbers of iterations for the (837,726) code with different
SNRs are listed in Table~\ref{tab:ave_conv_steps_837}. \CONV{The average numbers
  of iterations required by the IISRB and RS-IISRB algorithms are reduced by at least 20\% when the
$I_{\rm{max}}=50$. Iterations required by the RS-IISRB algorithm is slightly
fewer than that of the IISRB algorithm because of the re-selection scheme.}
\ALARM{In addition, we compare
the running time of different decoding algorithms (implemented in C) on a
DELL Optiplex 755. To decode 10,000 codewords of the (837,726) code over the AWGN
channel at the SNR of 5.4 dB, the ISRB, IISRB and RS-IISRB algorithms run
22.22, 19.48 and 19.37 seconds, respectively. In terms of the running time,
RS-IISRB $<$ IISRB $<$ ISRB, which is consistent with the comparison based on the average
number of iterations.}

\begin{table}[htbp]
\centering
\caption{Average number of iterations of the ISRB,
  IISRB and RS-IISRB algorithms for the (837, 726) code when $I_{\rm{max}}=50$
  and the modulation scheme is BPSK over the AWGN channel}
\label{tab:ave_conv_steps_837}
\begin{tabular}{|c|c||c|c|}
\hline
$E_b/N_0$ (dB) & ISRB \cite{Chaoyu2010mpa}& IISRB & RS-IISRB\\ \hline
4.5 & 22.18 & 10.58 & 9.97 \\ \hline
4.6 & 16.52 & 8.22 & 7.62 \\ \hline
4.7 & 12.69 & 6.54 & 6.30 \\ \hline
4.8 & 9.40  & 5.49 & 5.33 \\ \hline
4.9 & 7.44  & 4.79 & 4.64 \\ \hline
5.0 & 5.99  & 4.20 & 4.12 \\ \hline
5.1 & 5.21  & 3.80 & 3.71 \\ \hline
5.2 & 4.53  & 3.44 & 3.38 \\ \hline
5.3 & 4.06  & 3.15 & 3.10 \\ \hline
5.4 & 3.66  & 2.90 & 2.87 \\ \hline
\end{tabular}
\end{table}

For C1, C2 and C3, the BLERs of
the RS-IISRB algorithm are shown with the dashed curves in
Fig.~\ref{fig:per_ef_cw}. For C1 and C2, $\xi_1=4$, $\xi_2=1$, $\lambda=16$,
$\zeta=32$. For C3, $\xi_1=5$, $\xi_2=1$, $\lambda=16$,
$\zeta=32$. The RS-IISRB algorithm improves the BLER
performance and lowers the error floor for all three codes. In
Fig.~\ref{fig:per_ef_cw}, for C1, the simulation result for the 
RS-ISRB algorithm is shown as well, which does not adopt the new reliability information
update but the re-selection scheme. It appears that the re-selection scheme also provides
some performance gain. If both improvements are applied, the
RS-IISRB algorithm achieves a greater performance gain.

Next, we compare the performances of hard-reliability based MLGD algorithms. The
EIHRB-INIT algorithm \cite{Xinmiao2011EIHRB} is a
simplified version of the EIHRB algorithm without the recalculation of the
extrinsic information. The RS-IEIHRB algorithm is developed by integrating the
re-selection scheme describe in Section~\ref{subsec:RS} into the IEIHRB
algorithm.

\begin{figure}[htbp]
\centering
\includegraphics[width=9cm]{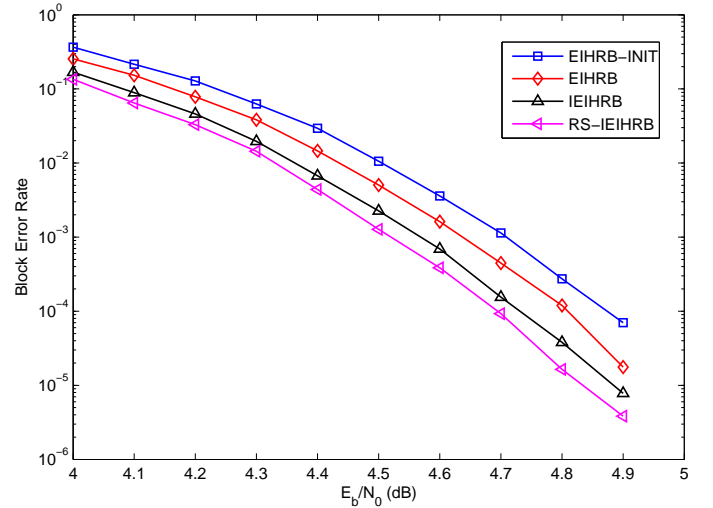}
\caption{Block error rates of hard-reliability based algorithms for the
  (255,175) code when $I_{\rm{max}}=50$ and the modulation scheme is BPSK over the AWGN channel}
\label{fig:BLER_HRB_255}
\end{figure}

Fig.~\ref{fig:BLER_HRB_255} shows the BLERs of different 
hard-reliability based algorithms for the (255,175) code, and Table~\ref{tab:ave_conv_steps_255_HRB}
lists the average numbers of iterations when
$I_{\rm{max}}=50$. For the EIHRB-INIT and EIHRB algorithm, $c_1=4$ and
$c_2=15$. For the IEIHRB and RS-IEIHRB algorithm, $c_1=1$, $c_2=63$,
$c_3=12$ and $\zeta=16$. For the (255,175) code the new reliability
information update provides about 0.05 dB performance gain, and the
re-selection scheme provides another 0.05 dB performance gain. Hence,
compared with the EIHRB algorithm, the RS-IEIHRB algorithm has about
0.1 dB performance gain, and \CONV{the average number of iterations required by the RS-IEIHRB algorithm is
reduced by about $30\%$.}

\begin{figure*}[htbp]
\centering
\includegraphics[width=14cm]{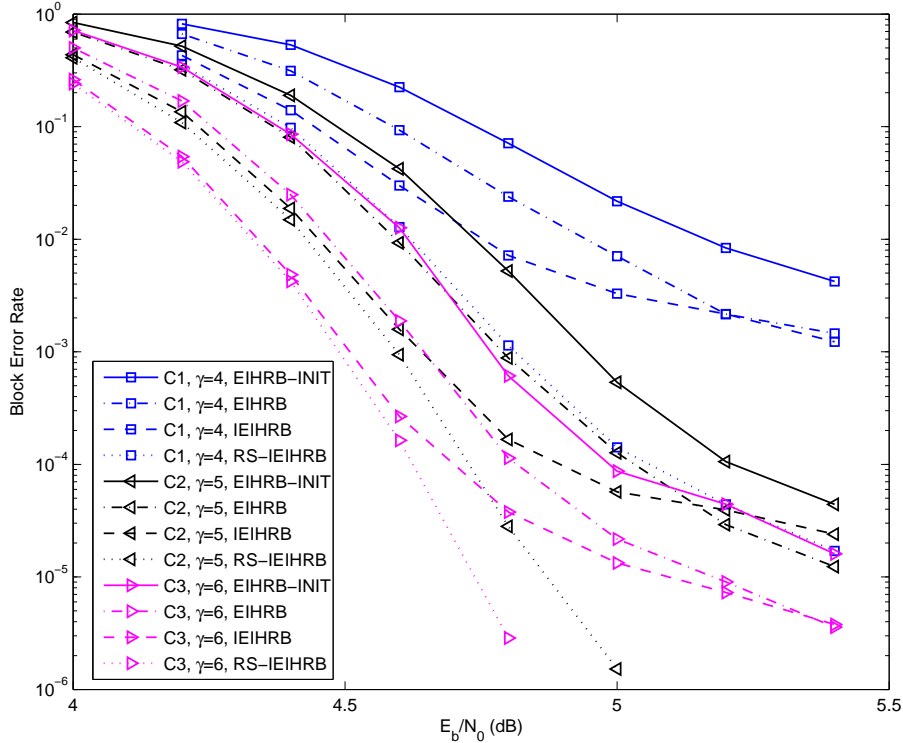}
\caption{Block error rates of hard-reliability based algorithms for different
  non-binary LDPC codes with different column weights when $I_{\rm{max}}=50$ and
  the modulation scheme is BPSK over the AWGN channel}
\label{fig:per_HRB_837}
\end{figure*}

\begin{table}[htbp]
\centering
\caption{Average number of iterations of the
  hard-reliability based algorithms for the (255, 175) code when
  $I_{\rm{max}}=50$ and the modulation scheme is BPSK over the AWGN channel}
\label{tab:ave_conv_steps_255_HRB}
\begin{tabular}{|c|c|c||c|c|}
\hline
$E_b/N_0$ & EIHRB-INIT  & EIHRB  & IEIHRB & RS-IEIHRB\\ 
(dB) & \cite{Xinmiao2011EIHRB} & \cite{Xinmiao2011EIHRB} & & \\ \hline
4.0 & 22.28 & 18.56 & 12.67 & 12.09 \\ \hline
4.1 & 16.48 & 13.62 & 8.99  & 8.85  \\ \hline
4.2 & 12.21 & 9.54  & 6.42  & 6.09  \\ \hline
4.3 & 8.57  & 7.20  & 4.70  & 4.62  \\ \hline
4.4 & 6.44  & 5.57  & 3.70  & 3.69  \\ \hline
4.5 & 5.15  & 4.70  & 3.10  & 3.10  \\ \hline
4.6 & 4.29  & 4.07  & 2.73  & 2.72  \\ \hline
4.7 & 3.75  & 3.65  & 2.47  & 2.47  \\ \hline
4.8 & 3.38  & 3.33  & 2.28  & 2.28  \\ \hline
4.9 & 3.09  & 3.06  & 2.14  & 2.14  \\ \hline
\end{tabular}
\end{table}

Fig.~\ref{fig:per_HRB_837} compares the BLERs of hard-reliability based
algorithms for different non-binary LDPC codes with different column
weights. For C1 and C2, $c_1=10$, $c_2=63$, $c_3=2$ and $\zeta=32$. For C3, $c_1=11$,
$c_2=63$, $c_3=2$ and $\zeta=32$. For the (837,726) code, the EIHRB algorithm also has an
error floor of $10^{-3}$. For low SNRs, the IEIHRB algorithm outperforms
the EIHRB algorithm and the RS-IEIHRB algorithm
reduces the error floor to a level of $10^{-5}$. In the error floor region, the EIHRB algorithm is
better than the IEIHRB algorithm because of the use of $z_{i,t}^{(k)}$ and
recalculating the extrinsic information in the latter. The two improvements in
Section~\ref{sec:algorithm} also \CONV{help to reduce the average number of iterations by about
$20\%$} for the (837,726) code as listed in Table~\ref{tab:ave_conv_steps_837_hd}. For C2 and C3, the new reliability
information update provides some performance gains for low SNRs, and the error
floors are lowered effectively.

\begin{table}[htbp]
\centering
\caption{Average number of iterations of the hard-reliability
  based algorithms algorithm for the (837, 726) code when $I_{\rm{max}}=50$ and
  the modulation scheme is BPSK over the AWGN channel}
\label{tab:ave_conv_steps_837_hd}
\begin{tabular}{|c|c|c||c|c|}
\hline
$E_b/N_0$ & EIHRB-INIT & EIHRB & IEIHRB & RS-IEIHRB\\
(dB) &  \cite{Xinmiao2011EIHRB} &  \cite{Xinmiao2011EIHRB} & & \\ \hline
4.2 & 43.01 & 37.25 & 28.83 & 27.49 \\ \hline
4.4 & 32.19 & 23.09 & 15.52 & 14.55 \\ \hline
4.6 & 18.85 & 12.41 & 8.29  & 7.83  \\ \hline
4.8 & 10.57 & 7.60  & 5.45  & 5.22  \\ \hline
5.0 & 6.79  & 5.49  & 4.13  & 4.01  \\ \hline
5.2 & 4.98  & 4.37  & 3.36  & 3.27  \\ \hline

\end{tabular}
\end{table}

%
\vspace{3mm}

\begin{figure}[htp]
\centering
\includegraphics[width=9cm]{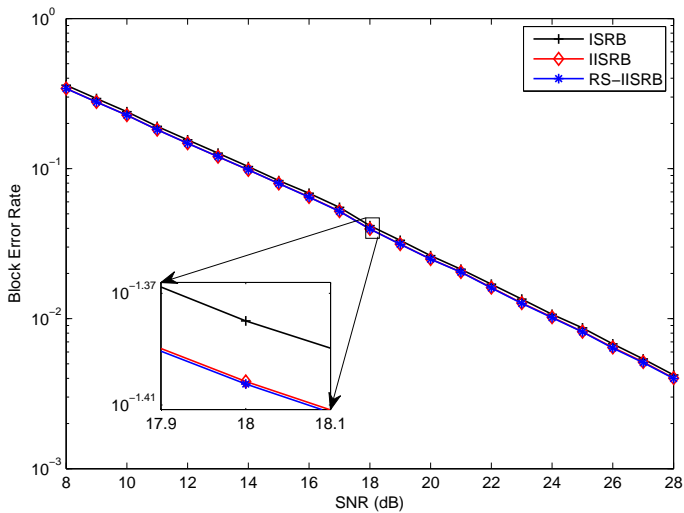}
\caption{Block error rates of algorithms for the (837, 726) code when
  $I_{\rm{max}}=50$ and the modulation scheme is BPSK
  over the block Rayleigh fading channel}
\label{fig:BRL_per}
\end{figure}

\begin{figure}[htp]
\centering
\includegraphics[width=9cm]{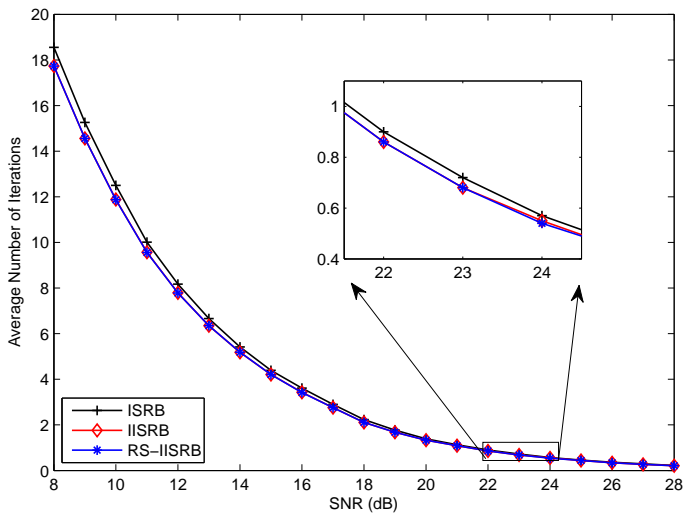}
\caption{Average numbers of iterations for different algorithms for the (837, 726) code when
  $I_{\rm{max}}=50$ and the modulation scheme is BPSK
  over the block Rayleigh fading channel}
\label{fig:BRL_iter}
\end{figure}

\ALARM{We evaluate the proposed decoding algorithms over block fading channels,
  which are widely used in wireless communication systems 
involving slow time-frequency hopping or multi-carrier modulation using
orthogonal frequency division multiplexing technique. We assume that each
codeword experiences a block Rayleigh fading channel and that the receiver has
perfect channel state information. Fig.~\ref{fig:BRL_per} and
Fig.~\ref{fig:BRL_iter} show the BLERs and the average numbers of iterations of
different MLGD algorithms for the (837,726) code over a block Rayleigh
fading channel. In Fig.~\ref{fig:BRL_per}, the IISRB and RS-IISRB algorithms have a gain
of about 0.2 dB over the ISRB algorithm, which is similar to that over the AWGN
channel shown in Fig.~\ref{fig:BLER_837}. At a SNR of 23 dB, the proposed
improvements reduce the average number of iterations by about $5\%$.}

In a word, the two improvements introduced in Section~\ref{sec:algorithm} apply to both the
soft-reliability and hard-reliability based MLGD algorithms. While both
improvements improve the error performance and \CONV{require fewer iterations on average}, the re-selection scheme lowers the error floor of codes
having low column weights effectively.

\subsection{Computational Complexity Reduction}
\label{sec:complexity}

We evaluate impacts on the complexity by the two proposed improvements and focus
on the soft-reliability based MLGD algorithms first. Assume the quantized
input information $q_{j,t}$ has a bit width of $\omega$. Without the clipping operation of
Eq.~\eqref{clipping}, for the ISRB algorithm, $R_{j,l}^{(k)}$ needs $\omega+\lceil
\log_2((\lambda+I_{\rm{max}}\gamma)r) \rceil$ bits and its bit width increases as $I_{\rm{max}}$ grows. However, for
the IISRB algorithm, $R_{j,l}^{(k)}$ needs only
$\omega+\lceil\log_2((\xi_1+\xi_2\gamma)r) \rceil$ bits and $I_{max}$ has no
impact on $R_{j,l}^{(k)}$'s bit width. With the clipping operation,
$R_{j,l}^{(k)}$ needs a smaller bit width in the ISRB algorithm. However, $N2^r$ IAs and $N(2^r-1)$ ICs
are needed per iteration to carry out the clipping operation. In contrast, there
is no accumulation operation in the IISRB and RS-IISRB algorithms. Thus, saturation is not an
issue for the IISRB and RS-IISRB algorithms, and the clipping operation is not needed.

For the IISRB algorithm, let us consider the initialization step first. 
There are $N2^r$ $\varphi_{j,l}$'s. To compute $\varphi_{j,l}$'s needs $Nr2^r$ IAs. Because $\max_l \varphi_{t,l} =
\varphi_{t,z_t}$ and there are $N\gamma$ $\phi_{i,j}$'s, $N\gamma(\rho-2)$ ICs
are needed to calculate $\phi_{i,j}$'s. 
The calculations of $\xi_1\varphi_{j,l}$'s and $\xi_2\phi_{i,j}$'s need
$N2^r$ and $N\gamma\rho$ IMs, respectively. Therefore, the initialization step
needs $Nr2^r$ IAs, $N2^r+N\gamma\rho$ IMs and $N\gamma(\rho-2)$ ICs. 

We now analyze the complexity per iteration of the IISRB algorithm.
Each iteration needs $M\rho$ FMs and $M(\rho-1)$ FAs to
calculate the syndrome $\mathbf{s}^{(k)}$. Line~\ref{r6} in Alg.~\ref{alg:ISRB} can
be reformulated as: 
\begin{equation}
\label{eq:sigma}
\sigma_{i,j}^{(k)}=h_{i,j}^{-1}s_i^{(k)}+z_j^{(k)}
\end{equation}
Hence, $N\gamma$ FAs and $N\gamma$ FMs are needed to calculate $\sigma_{i,j}^{(k)}$'s. Assume
there are $u_j^{(k)}$ $( 0< u_j^{(k)} \leq \gamma)$ different values among
$\sigma_{i,j}^{(k)}$'s for each $j$, then $u_j^{(k)}$ $R_{j,l}^{(k+1)}$'s
need to be updated. To compute $\psi_{j,l}^{(k)}$'s and $R_{j,l}^{(k+1)}$'s,
$\gamma-u_j^{(k)}$ and $u_j^{(k)}$ IAs are needed, respectively, for each
$j$. For $z_j^{(k+1)}$, $R_{j,z_j^{(k+1)}}^{(k+1)}$ must be
one of $R_{j,z_j}^{(k+1)}$ and those $R_{j,l}^{(k+1)}$ updated in the $k$-th iteration. To make the hard
decisions, $N\gamma$ ICs are needed at most. Hence, in the worst case,
each iteration of the IISRB algorithm requires $2N\gamma$ FMs, $2N\gamma-M$ FAs, $N\gamma$ IAs and
$N\gamma$ ICs $(M\rho=N\gamma)$. Compared with the ISRB
algorithm, the IISRB algorithm saves $N2^r$ IAs and $N(2^{r+1}-2-\gamma)$ ICs for each iteration,
while requiring the same numbers of FAs and FMs. This saving is significant if
$2^r$ is large.

Let us calculate computational complexity overhead due to the
re-selection scheme. 
$\tilde{\mathbf{z}}=(\tilde{z}_0, \tilde{z}_1,
\cdots, \tilde{z}_{N-1})$ represents the second most likely decision of
the received word $\mathbf{y}$. To acquire $\tilde{z}_j$ in the initialization
step, $r-1$ ICs are needed for each $j$, because $r$-bit representations of 
$\tilde{z}_j$ and $z_j$ differ by one bit and there are $r$ elements over
GF($2^r$) satisfying this constraint. Hence, the initialization step of the
RS-IISRB algorithm needs $N(r-1)$ ICs more than that of the IISRB algorithm. 
For each iteration, the second maximum among $R_{j,l}^{(k)}$'s must
be one of $R_{j,\tilde{z}_j}^{(k)}$, $R_{j,z_j}^{(k)}$ and those
$R_{j,l}^{(k)}$'s updated. It needs at most $N(\gamma+1)$ ICs per iteration. Line
\ref{r7} of Alg.~\ref{alg:RS-Scheme} needs $2N$ ICs to identify 
the existence of a periodic point. $N\gamma$ IAs and $N\gamma$ ICs are needed to 
calculate $\text{us\_c}_j$. The calculation of $\rm{dif\_R}$ needs $2N$ ICs and $N$
IAs. $\varphi_{{\rm{rs\_n}},z_{{\rm{rs\_n}}}^{(k)}}$ and 
$\varphi_{{\rm{rs\_n}},\tilde{z}_{{\rm{rs\_n}}}^{(k)}}$ need two IAs. After the
re-selection scheme, there are $\gamma$ syndromes to be recalculated so that
Eq.~\eqref{eq:sigma} can be applied, requiring $2\gamma$ FAs and $\gamma$
FMs. Therefore, the RS-IISRB algorithm needs $2\gamma$ FAs, $\gamma$ FMs,
$N\gamma+N+2$ IAs and $5N+2N\gamma$ ICs per iteration more than the IISRB
algorithm. 
\begin{table*}[htbp]
\centering
\caption{Computational complexities of the initialization step for various
decoding algorithm to decode an LDPC code over GF($2^r$) with an $M\times N$
parity check matrix whose column and row weights are $\gamma$ and $\rho$}
\label{tab:complexity_table_general_init}
\begin{tabular}{|c|c|c|c|c|c|}
\hline
Algorithms & IA & IM & IC & ID & Floor\\ \hline
ISRB \cite{Chaoyu2010mpa} & $Nr2^r$ & $N2^r$ & $MN(2^r-1)(3\rho-6)$ & 0 & 0\\ \hline
IISRB & $Nr2^r$ & $N2^r+N\gamma\rho$ & $N\gamma(\rho-2)$ & 0 & 0 \\ \hline
RS-IISRB &  $Nr2^r$ & $N2^r+N\gamma\rho$ & $N\gamma(\rho-2)+N(r-1)$ & 0 & 0\\ \hline
\hline
EIHRB \cite{Xinmiao2011EIHRB} & $N2^r(r+2)$ & 0 & $N2^r$ & $N2^r$ & $N2^r$ \\ \hline
IEIHRB & $N2^r(r+2)$ & 0 & $N2^r$ & $N2^r$ & $N2^r$ \\ \hline
RS-IEIHRB & $N2^r(r+2)$ & 0 & $N2^r+N(r-1)$ & $N2^r$ & $N2^r$ \\ \hline
\end{tabular}
\end{table*}

\begin{table*}[htbp]
\centering
\caption{Computational complexities per iteration for various
decoding algorithm to decode an LDPC code over GF($2^r$) with an $M\times N$
parity check matrix whose column and row weights are $\gamma$ and $\rho$}
\label{tab:complexity_table_general}
\begin{tabular}{|c|c|c|c|c|}
\hline
Algorithms & FA & FM& IA & IC \\ \hline
ISRB \cite{Chaoyu2010mpa} & $2N\gamma-M$ & $2N\gamma$ & $N\gamma+N2^r$ & $2N2^r-2N$ \\ \hline
IISRB & $2N\gamma-M$ & $2N\gamma$ & $N\gamma$ & $N\gamma$ \\ \hline
RS-IISRB & $2N\gamma-M+2\gamma$ & $2N\gamma+\gamma$ & $2N\gamma+N+2$ & $5N+3N\gamma$ \\ \hline
\hline
EIHRB \cite{Xinmiao2011EIHRB}& $3N\gamma-2M$ & $3N\gamma$ & $2N\gamma+N2^r$ & $2N2^r-2N+N\gamma$ \\ \hline
IEIHRB & $2N\gamma-M$ & $2N\gamma$ & $N\gamma$ & $N\gamma$ \\ \hline
RS-IEIHRB & $2N\gamma-M+2\gamma$ & $2N\gamma+\gamma$ & $2N\gamma+N+2$ & $5N+3N\gamma$ \\ \hline
\end{tabular}
\end{table*}

Complexities of the hard-reliability based algorithms can be analyzed
similarly. The IEIHRB algorithm has the same computational complexity per
iteration as the IISRB algorithm, because they have the same iteration
procedure. For the same reason, the RS-IEIHRB algorithm has the same computational complexity per
iteration as the RS-IISRB algorithm.

Tables~\ref{tab:complexity_table_general_init} and
\ref{tab:complexity_table_general} compare computational complexities of
various decoding algorithms. For the initialization step, the numbers of IMs of the IISRB and RS-IISRB
algorithms are greater than that of the ISRB algorithm because the calculation
of $\xi_2\phi_{i,j}$ is done in initialization to reduce computational complexities of iterations. This is a
good trade-off for computational complexity. The number of ICs needed by the initialization step
of the ISRB algorithm provided in \cite[Section III-A]{Chaoyu2010mpa} is
significantly greater than those of the other algorithms. This is because in
\cite{Chaoyu2010mpa}, $\phi_{i,j}$'s are calculated for every $i$ and $j$, and
$\max_l \varphi_{t,l}$'s are re-calculated for each $\phi_{i,j}$.
For each iteration, the numbers of integer operations
required by the ISRB and EIHRB algorithms scale with $2^r$, the
order of the finite field. With
the new reliability information update, the numbers of integer operations are reduced greatly
and are now independent of $2^r$. The re-selection scheme incurs some additional
complexity, but complexities of the RS-IISRB and RS-IEIHRB algorithms are still lower than
those of the ISRB and EIHRB algorithms, respectively.

Tables~\ref{tab:complexity_table_specific_init} and
\ref{tab:complexity_table_specific} list the numbers of
various operations for initialization and each iteration, respectively, needed by various decoding algorithms for
the (255,175) code. For initialization, the ISRB
algorithm needs significantly more ICs than the other algorithms. When the order of the finite
field is higher, our improved algorithms reduce the numbers of IAs and ICs for
each iteration significantly. 

From a perspective of the computational complexity, the IISRB
and IEIHRB algorithms are the best. The re-selection scheme needs more finite field operations and integer
operations. All the improved algorithms are simpler than the
ISRB and EIHRB algorithms. 

Let us consider the memory overhead required by the two improvements. Our first improvement---the new reliability information update---does
not need any extra memory units. The second improvement---the re-selection scheme---needs to store $\mathbf{z}^{(k-1)}$ and
$\mathbf{z}^{(k-2)}$ and hence requires $2Nr$ extra memory bits. 
Hence, the re-selection scheme increases the memory requirement slightly, but it does lower the error floor.

\begin{table}[!htp]
\centering
\caption{Computational complexities of the initialization step for various
decoding algorithms to decode the (255,175) code}
\label{tab:complexity_table_specific_init}
\begin{tabular}{|c|c|c|c|c|c|}
\hline
Algorithm & IA & IM & IC & ID &Floor \\ \hline
ISRB \cite{Chaoyu2010mpa} & 522240 & 65280 & 696417750 & 0& 0 \\ \hline
IISRB & 522240 & 130560 & 57120 & 0 & 0 \\ \hline
RS-IISRB & 522240 & 130560 & 58905 & 0 & 0 \\ \hline
\hline
EIHRB \cite{Xinmiao2011EIHRB} & 522750 & 0 & 65280 & 65280 & 65280 \\ \hline
IEIHRB & 522750 & 0 & 65280 & 65280 & 65280 \\ \hline
RS-IEIHRB & 522750 & 0 & 67065 & 65280 & 65280 \\ \hline
\end{tabular}
\end{table}

\begin{table}[!htp]
\centering
\caption{Computational complexities required per iteration for various
decoding algorithm to decode the (255,175) code}
\label{tab:complexity_table_specific}
\begin{tabular}{|c|c|c|c|c|}
\hline
Algorithm & FA & FM& IA & IC \\ \hline
ISRB \cite{Chaoyu2010mpa} & 7905 & 8160 & 69360 & 130050 \\ \hline
IISRB & 7905 & 8160 & 4080 & 4080 \\ \hline
RS-IISRB & 7937 & 8176 & 8417 & 13515 \\ \hline
\hline
EIHRB \cite{Xinmiao2011EIHRB} & 11730 & 12240 & 73440 & 134130 \\ \hline
IEIHRB & 7905 & 8160 & 4080 & 4080 \\ \hline
RS-IEIHRB & 7937 & 8176 & 8417 & 13515 \\ \hline
\end{tabular}
\end{table}

\section{Conclusion}
\label{sec:conclusion}

In this paper, we propose two improvements to the soft-reliability and
hard-reliability based MLGD algorithms for non-binary LDPC codes. The first
improvement---the new
reliability information update---helps the reliability-based MLGD algorithms
achieve better BLERs, \CONV{require fewer iterations}, and have lower complexities. The second
improvement---the re-selection scheme---results in a better error
performance, \CONV{fewer iterations on average}, and a lower error floor. Although the
re-selection scheme needs additional complexity, the MLGD algorithms with
the re-selection scheme still require lower computational complexities than the
existing MLGD algorithms.

%
%




%

\end{document}